\documentclass[12pt]{article}
\usepackage{amssymb}
\usepackage{amsmath}
\usepackage{hyperref}
\usepackage{graphicx}
\usepackage{placeins}
\usepackage{bm}
\usepackage{latexsym}
\usepackage[font=small,labelfont=bf]{caption}
\begin{document}
\title{\bf{Gravitational reduction of the wave function  through the quantum theory of motion}}
\author{\underline{Faramarz Rahmani} \thanks{Corresponding author: Email: farmarz.rahmani@abru.ac.ir
}  \\
		 {\small {\em  *Department of Physics, Faculty of Basic Sciences, Ayatollah Boroujerdi University, Boroujerd, Iran}}
}				
   
\date{\today}
\maketitle

\abstract{We present a novel perspective on gravity-induced wave function reduction using Bohmian trajectories. This study examines the quantum motion of both point particles and objects, identifying critical parameters for the transition from quantum to classical regimes. By analyzing the system's dynamics, we define the reduction time of the wave function through Bohmian trajectories, introducing a fresh viewpoint in this field. Our findings align with results obtained in standard quantum mechanics, confirming the validity of this approach.}



\section{Introduction} \label{intro}
The reduction of the wave function, a long-standing unresolved issue in quantum physics, has been studied through various approaches. For instance, GRW theory suggests that wave function collapse occurs spontaneously and randomly over time \cite{Ghirardi:1989bb, g2}. Decoherence theory explains how interactions with the environment cause a quantum system to lose coherence, thereby behaving in a classical manner \cite{d1, Giulini:1996nw, Zeh:1995jg}. The many-worlds interpretation posits that all possible outcomes of a measurement actually occur, each in its own separate branch of the universe, resulting in a vast multiverse where every possible result is realized in some branch \cite{Saunders:2012zz, Schlosshauer:2022ycx}.
In our study, we explore the gravitational reduction of the wave function. This approach has been extensively examined within the context of standard quantum mechanics (see Ref.\cite{RefBassi}). However, we aim to investigate this topic through the lens of Bohmian quantum mechanics, which offers greater intuition and insight into the underlying processes.\par 
In gravity-induced wave function reduction, the mass of an object serves as a criterion for determining the boundary between the classical and quantum worlds. In this interpretation, the self-gravity of the object or the curvature of spacetime caused by its mass directs the object to a specific state. This concept, first formulated around sixty years ago \cite{RefK1,RefK2,RefK3}, remains relevant. Macroscopic objects adhere to the laws of classical mechanics, while the behavior of microscopic objects is governed by the Schr\"{o}dinger equation. Thus, defining a critical mass is essential to demarcate the boundary between the classical and quantum realms. \par 
The concept of self-gravity should not be confusing. Consider a particle or object to which a wave packet is attributed. According to the principles of quantum mechanics, this particle or object can be in multiple places simultaneously, meaning we have a superposition of different states of the particle. Within this probability space, a distribution of matter locations can be considered with self-gravitational energy. Since this distribution is rooted in the uncertainty principle, we refer to its associated energy as 'quantum gravitational energy.' The quantum superposition is unstable in the presence of such a gravitational field and is disrupted when the mass of the particle or object reaches the critical mass. When the superposition breaks down, the particle or object moves towards greater certainty. The timescale between breaking the superposition and the system settling into a specific state is often referred to as the reduction time of the wave function. This concept aligns closely with the ideas of Penrose and Karolyhazi, while the damping time proposed by Diosi is also relevant to this concept.
This topic has been extensively studied in standard quantum mechanics. Now, we will examine it within the framework of Bohmian quantum mechanics, where definitions such as trajectory, force, and velocity are meaningful. Our goal is to provide a fresh perspective on the concept of reduction time. Despite numerous achievements in standard quantum mechanics, an ontological and deterministic description of the problem remains elusive. We cannot predict with certainty which of the possible states the quantum system will occupy after measurement or the breaking of superposition by gravity, and we rely solely on statistical predictions. This unpredictability persists in Bohmian quantum mechanics as well.\par 
First, let's explore the gravitational approaches that exist in standard quantum mechanics. The first significant work in this area is by Karolyhazi, who considered an uncertainty for the metric of spacetime. To achieve the classical regime, the uncertainty in spacetime is minimized in the presence of a scalar field. This leads to the critical conditions necessary for the transition from the quantum domain to the classical domain. Karolyhazy's relations for the critical width and reduction time of an elementary particle are given by
\begin{equation}\label{karol1}
\sigma_c = \frac{\hbar^2}{Gm^3}
\end{equation}
and
\begin{equation}\label{karol2}
\tau=\frac{m \sigma_c^2}{\hbar}.
\end{equation}
Here, the reduction time is the time it takes for a quantum state to undergo spontaneous localization, leading to a classical-like behavior. For a proton, one can find $\sigma_c\approx 10^{22} m$ and $\tau  \approx 10^{53} s$. For an object of mass $M$ and size $R$, Karolyhazy derives the following relation:
\begin{equation}\label{karol3}
\sigma_c \sim \Big(\frac{\hbar^2}{Gm^3}\Big)^{\frac{1}{3}} R^{\frac{2}{3}}.
\end{equation}
This relation helps in determining the critical conditions necessary for the transition from the quantum domain to the classical domain based on the object's mass and size.
For a tennis ball of radius $\sim 4 cm$ and mass $\sim 57 g$, the critical width and reduction time are approximately $\sigma_c\sim 10^{-17}cm$ and $\tau \sim 10^{-6}s$ respectively. For details, see\cite{RefK1,RefK2,RefK3}.\par 
The gravitational approach of Diosi also determines a critical width for the transition from the quantum to the classical world. Diosi's first significant work is an investigation through the Schrödinger-Newton equation, a nonlinear equation with a self-gravitational potential. It is believed that such a nonlinear term breaks down the unitary transformation and transforms a pure state into a mixture of states. By using the Schrödinger-Newton equation, the critical width for the transition is obtained.
Subsequent works by Diosi deal with the concept of intrinsic quantum imprecision in the spacetime metric. By studying the fluctuations of the gravitational field when measured by a quantum probe over a period of time, the damping time or reduction time of a quantum state in the presence of gravity is defined \cite{RefD1, RefD2, RefD3}. The critical widths for the transition from the quantum world to the classical world of an object in Diosi's approach are as follows:
\begin{equation}\label{dio11}
\sigma_c = \Big(\frac{\hbar^2}{Gm^3}\Big)^{\frac{1}{4}} R^{\frac{3}{4}}, \quad \text{macro objects}
\end{equation}
and
\begin{equation}\label{dio12}
\sigma_c = \Big(\frac{\hbar^2}{Gm^3}\Big)^{\frac{1}{2}} R^{\frac{1}{2}}, \quad \text{micro objects}.
\end{equation}
In the above relation, a macro object refers to a macroscopic  body that exhibits classical behavior and does not exhibit significant quantum superposition over macroscopic distances. A micro object exhibits quantum behaviors such as superposition and entanglement. We expect that a micro object to be in different places or states at the same time and the width of its wave packet to be greater than its classical size. For a proton, the critical width is about $10^{4} m$, which is really greater than its classical size,.i.e $10^{-15} m$. The damping time or the reduction time for a proton is about $10^{15}s\approx 32000000 $ years, which indicates that we can not see the collapse of the proton normally.\par 
The approach of Penrose is very fundamental. In Ref. \cite{RefP1}, an effort is made to bring quantum mechanics closer to the principles of general relativity. Penrose shows that in order to establish the principles of general relativity at the quantum level, a quantum superposition must be broken and the system in question must collapse into one of the possible states. In other words, Penrose proposes that a quantum state remains in superposition until the difference in spacetime curvature reaches a significant level.
Penrose suggests that the wave function cannot sustain superposition beyond a certain energy difference between quantum states. When the energy difference exceeds a threshold, the wave function collapses to a single state based on its original amplitude. An interesting point is that observers do not play a special role in the collapse process. According to these statements, macroscopic systems cannot exist in multiple places simultaneously due to the large energy difference. Microscopic systems (such as electrons) can remain in superposition for longer (thousands of years) until their spacetime curvature separation reaches the collapse threshold.
In Penrose's approach, an ensemble of spacetimes is inevitably considered. Since the superposition of spacetimes is illegal and a well-defined timelike Killing vector cannot be defined in this situation, the system must collapse into a particular state.
\par  
Penrose highlights another intriguing conceptual point. He argues that to establish the principle of equivalence at the quantum level, a unitary transformation must relate the wave functions of an accelerated observer (Newtonian observer) and an inertial observer (Einsteinian observer). The nonlinear dependence of the phase of this transformation on time causes the concept of 'positive frequency' to differ for the two observers, resulting in them experiencing different spacetimes. The aforementioned unitary transformation is as follows:
\begin{equation}\label{w}
 \psi(\mathbf{x}^{\prime},t^{\prime})=\phi(\mathbf{x},t)\exp \Big(  \frac{i}{\hbar}(m\mathbf{g}\cdot \mathbf{x}^{\prime} t^{\prime}+\frac{1}{3}m\mathbf{g}^2 t^{\prime 3})\Big).
\end{equation} 
Where, $\psi(\mathbf{x}^{\prime},t)$ and $\phi(\mathbf{x},t)$ are the wave functions of the Newtonian and Einsteinian observers respectively. 
This is the same physical concept we are familiar with in the theory of quantum fields in curved spacetime, during which an accelerated observer does not see the quantum field as a pure field but as a mixture of states, i.e., a statistical system with a particular temperature. This particular temperature is called the Unruh temperature. During wave function reduction, a pure state also transforms into a mixture of states. Therefore, a concept similar to the Unruh temperature may also be derived from this phenomenon. These concepts are discussed in Refs. \cite{RefP2, RefP3}. In Ref. \cite{RefRR1}, we have shown that it is possible to define a temperature that closely resembles the Unruh temperature in the problem of wave function reduction. To study the gravitational reduction of the wave function within the framework of standard quantum mechanics, Refs. \cite{RefGu, Refm1, Refde, Howl:2018qdl, Tagg:2024fvq, RefBassi2} are very suitable.
\par 
Here, it is appropriate to mention the general features of Bohmian quantum mechanics. Bohmian quantum mechanics is an approach in which the particle has a deterministic quantum dynamics, allowing us to consider trajectories for the particles. In non-relativistic Bohmian quantum mechanics, a particle moves along a well-defined trajectory in the ensemble with the velocity
\begin{equation}\label{vel1}
\frac{d\mathbf{x}(t)}{dt}=\left(\frac{\nabla S(\mathbf{x},t)}{m}\right)_{\mathbf{X}=\mathbf{x}(t)}=\left(\frac{\mathbf{p}}{m}\right)_{\mathbf{x}=\mathbf{x}(t)},
\end{equation}
where $\mathbf{p}=\nabla S(x,t)$ is the momentum of the particle. The meaning of $\mathbf{x}=\mathbf{x}(t)$ is that the particle moves along a trajectory of the ensemble which is described by $\mathbf{x}(t)$. Here,  $S(x,t)$ is the principal function of the particle. 
Newton's second law for the quantum motion of the particle in an external potential, $U$, takes the form
\begin{equation}\label{dn}
\frac{d}{dt}(m\dot{\mathbf{x}})=-\nabla U-\nabla Q \vert_{\mathbf{x}=\mathbf{x}(t)}.
\end{equation}
Where, $-\nabla Q$, is the quantum force exerts on the particle and $Q$ denotes the quantum potential. The non-relativistic quantum potential for a spinless particle is as follows:
\begin{equation}\label{Q1}
Q=-\frac{\hbar^2}{2m}\frac{\nabla^2 \mathcal{R}}{\mathcal{R}}.
\end{equation}
where $\mathcal{R}$ is the amplitude of the wave function which can be represented in the following polar form 
\begin{equation}
\psi(\mathbf{x},t)=\mathcal{R} \exp \Big(\frac{iS(\mathbf{x},t)}{\hbar}\Big).
\end{equation}
The energy of the particle is given by
\begin{equation}
E=-\frac{\partial S}{\partial t},
\end{equation}
which is related to the modified Hamilton-Jacobi equation as follows:
\begin{equation}\label{hamilton}
\frac{\partial S(\mathbf{x},t)}{\partial t}+\frac{(\nabla S)^2}{2m}+U+Q=0.
\end{equation}
where $U$ is the classical potential energy. The quantum potential has non-local property and is responsible for the quantum behavior of the particle\cite{RefB,RefI,RefU,RefH,RefCush,RefCB}. For example, the dispersion of a wave packet can be related to the quantum force, which is the derivative of the quantum potential. In the famous two-slit experiment, it can be shown that the trajectories of particles passing through two slits are affected by the quantum force in such a way that the places where the particles collide on the screen form interference fringes.\par 
In Bohmian quantum mechanics, it is possible to study the quantum motion of the particle or object, which allows for a more intuitive visualization of the problem. We will see how concepts like trajectory and force help us easily obtain the critical quantities for the transition between the quantum and classical worlds. The reduction time of the wave function and its critical width will be obtained using the quantum dynamics of the particle or object. In previous studies, we considered the matter for a point particle \cite{RefRGG1, RefRGG2, RefRGG3}. However, in this paper, we will generalize it to an object, as what transitions from the quantum world to the classical world is no longer an elementary particle. In the framework of Bohmian quantum mechanics, there is no sharp distinction between the quantum and classical worlds. Instead, there is a relative interplay between the quantum force and gravity. The gravitational force characterizes the classical world, while the quantum force is intrinsic to the quantum world, with its origin still not fully understood. When the gravitational force dominates, the behavior of the particle or object exhibits less uncertainty. Conversely, when the quantum force is dominant, uncertainty becomes more prominent. In other words, the quantum force is not exactly zero in the classical world, just as the gravitational force is not zero in the quantum world.
\section{The Quantum Motion and Transition Condition}
\label{sec:2}
\begin{figure}[h!] 
\centerline{\includegraphics[width=6cm]{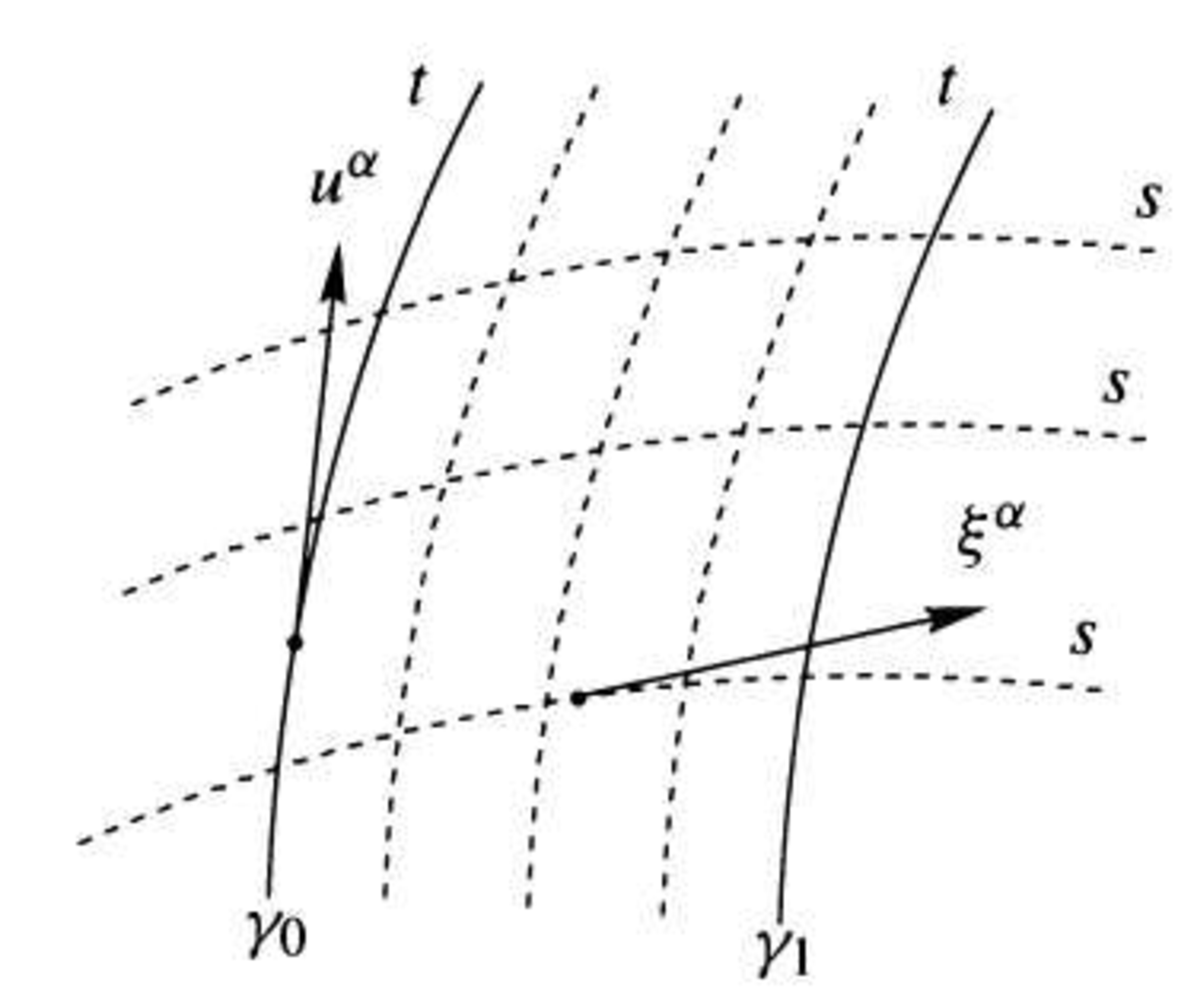}}
\caption{An ensemble of non-crossing trajectories. The deviation vector between two nearby trajectories is represented by  $\xi^{\alpha}$. The particle moves along one of the possible trajectories in the ensemble with the velocity $u^\alpha$. \label{fig:1}}
\end{figure}

In this section we provide a geometric approach for gravity-induced wave function reduction using Bohmian trajectories. We have done this before for a point particle. Now let's take a quick look at this approach and use the results for an object. It can be shown that the Bohmian trajectories do not cross each other\cite{RefH}. Thus, we can consider an ensemble of non-crossing trajectories in configuration space where an arbitrary trajectory $\gamma$ is parametrized by the parameter $t$ and the particle can choose one of these trajectories and move along it with the velocity vector $\mathbf{u}^{\alpha}$. See Fig.\ref{fig:1}. 
It should be noted that the usual gravity-induced wave function reduction models are effective models. So, we will consider a fixed background metric or gravitational field. The relative acceleration between two nearby trajectories is given by
\begin{equation}\label{adev}
\frac{\partial^2 \xi^{\alpha}}{\partial t^2}
= \frac{\partial ^2}{\partial t^2}\left(\frac{\partial x^{\alpha}(t,s)}{\partial s}\right)=\frac{\partial}{\partial s}\left( \frac{\partial^2 x^{\alpha}(t,s)}{\partial t^2}\right)=\frac{\partial a^{\alpha}}{\partial s}
\end{equation}
Where, $a^{\alpha}$ denotes the total acceleration of the particle. 
The involved forces are the usual quantum force and the gravitational force due to the quantum distribution ( $\varrho=\psi^* \psi$ ) in the configuration space of the particle which, we call the "quantum gravitational force". The quantum gravitational force(QGF) is the derivative of the quantum gravitational energy(QGE) which is defined as follows:
\begin{equation}\label{self}
U_{\text{QG}}(\mathbf{x},t)=\int \varrho(\mathbf{x}^\prime,t) U_{\text{cl}} d^3 \mathbf{x}^\prime =\int \vert \psi(\mathbf{x}^\prime,t)\vert^2 U_{\text{cl}} d^3 \mathbf{x}^\prime.
\end{equation}
Here, $U_{\text{cl}}$ denotes the classical potential energy of the particle or object. As mentioned in the introduction, the quantum gravitational energy refers to the gravitational energy of the quantum distribution of the particle or object in configuration space.\par 
Along the curves parametrized by $s$, we use $\frac{\partial}{\partial s} = \xi^{\beta} \frac{\partial}{\partial x^{\beta}}$ which helps us to write relation (\ref{adev}) in the form:
\begin{equation}\label{eq1}
\frac{\partial^2 \xi^{\alpha}}{\partial t^2}=-\xi^{\beta} \frac{\partial}{\partial x^{\beta}} \left(\frac{1}{m}\frac{\partial Q}{\partial x^{\alpha}}+\frac{1}{m}\frac{\partial U_{QG}}{\partial x^{\alpha}}\right).
\end{equation}
The terms in the parenthesis are quantum acceleration and quantum gravitational acceleration, respectively. In a gravity-dominant regime, where the quantum gravitational force overcomes the quantum force, we expect the trajectories to be closer together. Conversely, in a quantum-dominant regime, where the quantum force overcomes the quantum gravitational force, the trajectories of the ensemble diverge \cite{RefRGG3}.
It is obvious that to have zero deviation acceleration or parallel trajectories, we must have
\begin{equation}\label{eq2}
\frac{\partial U_{\text{QG}}}{\partial x^{\alpha}}=- \frac{\partial Q}{\partial x^{\alpha}}, \qquad \alpha=1,2,3.
\end{equation}
This can be written as:
\begin{equation}\label{eq}
\nabla Q=-\nabla U_{\text{QG}}.
\end{equation}
This relation will be used to obtain a criterion for the transition from the quantum world to the classical world. The condition $\nabla Q=-\nabla U_{\text{QG}}$ defines the transition regime in this context. We shall see that the average of the relation (\ref{eq}) gives the critical width of the wave packet for the transition between the quantum and classical domains. 
\subsection{The Transition Condition for a Point Particle}
Through the relation (\ref{eq1}), we obtain the condition (\ref{eq}), which states that when the quantum force and the quantum gravitational forces are balanced, any two nearby trajectories in the ensemble remain parallel. In this situation, we expect that the associated wave packet does not spread and remains a stationary wave packet. Thus, the average of relation (\ref{eq}) may lead us to the critical values for transition. \par 
We assume the particle is guided by a Gaussian wave packet. If we consider a homogeneous gravitational field, the form of the wave packet remains Gaussian. This is possible when we consider the problem in a short-time estimation \cite{RefH, RefRGG3}. We utilize the spherical form of the amplitude of the Gaussian wave packet for the sake of simplicity, which has the following form:
\begin{equation}\label{gauss1}
\mathcal{R}(r)= (2\pi \sigma^2)^{-\frac{3}{4}}e^{-\frac{r^2}{4\sigma^2}}.
\end{equation}
Here, $\sigma$ is the width of the wave packet at time $t$ which is given by
\begin{equation}
\sigma=\sigma_0 \sqrt{1+\frac{\hbar^2 t^2}{4m^2 \sigma_0^4}},
\end{equation}
where, $\sigma_0$ is the initial width of the wave packet. By short-time estimation, we mean $\sigma \approx \sigma_0$.
We shall see that the averages of gravitational force and gravitational potential depend on the width of the wave packet. Thus, in the short-time estimation, these quantities can be considered approximately constant.
Now, we evaluate the average of equation (\ref{eq}) to find a criterion for the transition regime. We first derive the relations of the quantum potential and the quantum gravitational potential of the particle. The quantum potential is obtained as follows:
\begin{equation}\label{sq}
Q=-\frac{\hbar^2}{2m}\frac{\nabla^2 \mathcal{R}}{\mathcal{R}}= -\frac{\hbar^2}{2m}\frac{1}{r}\frac{\partial}{\partial r}\left(r\frac{\partial R}{\partial r}\right)=\frac{\hbar^2}{8m\sigma_0^4}\left(6\sigma_0^2-r^2\right)
\end{equation}
The quantum gravitational potential is given by
\begin{equation}\label{pot2}
U_{\text{QG}}(r)= \int_{0}^{r} U_{\text{cl}}\varrho(r^{\prime}) 4\pi r^{\prime^2} dr^{\prime}=- \int_{0}^{r} \frac{Gm^2}{r^{\prime}}\varrho(r^{\prime}) 4\pi r^{\prime^2} dr^{\prime}=\sqrt{\frac{2}{\pi}}\frac{Gm^2}{\sigma_0}\left(1-e^{-\frac{r^2}{2\sigma_0^2}}\right).
\end{equation}
Where, $U_{\text{cl}}$ denotes the classical kernel of the system. Now, the quantum force and the quantum gravitational force are given by
\begin{equation}\label{sf}
f_{\text{Q}} =-\frac{\partial Q}{\partial r}=\frac{\hbar^2}{4m\sigma_0^4} r
\end{equation}
and 
\begin{equation}\label{sg}
f_{\text{QG}}=-\frac{\partial U_{\text{QG}}(r)}{\partial r}=-\sqrt{\frac{2}{\pi}}\frac{Gm^2}{\sigma_0^3}r e^{-\frac{r^2}{2\sigma_0^2}}.
\end{equation}
respectively. 
The average values of these quantities are 
\begin{equation}\label{asq}
\overline{f_{\text{Q}}} = \int_{0}^{\infty} \varrho(r) f_{\text{Q}} 4\pi r^2 dr=\frac{1}{2}\sqrt{\frac{2}{\pi}} \frac{\hbar^2}{m\sigma_0^3} 
\end{equation}
and
\begin{equation}\label{asf}
 \overline{f_{\text{QG}}}  = \int_{0}^{\infty} \varrho(r) f_{\text{QG}} 4\pi r^2 dr=-\frac{1}{\pi}\frac{Gm^2}{\sigma_0^2}
\end{equation}
respectively. By equating these results, we obtain the characteristic width for which the transition from the quantum to the classical world occurs. The transition width is as follows:
\begin{equation}\label{d1}
\sigma_0^{(c)}=\sqrt{\frac{\pi}{2}} \frac{\hbar^2}{G m^3} \approx \frac{\hbar^2}{G m^3}.
\end{equation}
By using above relation, the critical mass is defined as follows:
\begin{equation}\label{mc}
m_c=\left(\frac{\pi}{2}\right)^{\frac{1}{6}}\left(\frac{\hbar^2}{G \sigma_0}\right)^{\frac{1}{3}}\approx \left(\frac{\hbar^2}{G \sigma_0}\right)^{\frac{1}{3}}.
\end{equation}
This formula can be interpreted such that for a specific width 
$\sigma_0$, if the particle mass is greater than the critical mass, the quantum gravitational force overcomes the quantum force. Conversely, if the particle mass is less than the critical mass, the quantum force overcomes the quantum gravitational force, and the particle behaves in a quantum manner. This means its wave packet spreads with time and never reduces, leading to an increase in the uncertainty of the particle's position.
For example, consider a wave  packet with the initial width $\sigma_0=10^{-2} cm$. The critical mass for this wave packet is approximately $m_c\approx 10^{-15}g$. In other words, for this wave packet, this value of mass can balance the quantum gravitational and quantum forces so that the wave packet does not spread.  
By combination of relations (\ref{asq}), (\ref{asf}) and (\ref{mc}), we derive the following relation:
\begin{equation}\label{cri}
\frac{m_c}{m}=\left(\frac{\overline{f_{Q}}}{\overline{f_{\text{QG}}}}\right)^{\frac{1}{3}}.
\end{equation}
Through this relation, we conclude that:
\begin{equation}
\left\{ \begin {array} {cc} m>m_c \Leftrightarrow \overline{f_{Q}}<\overline{f_{QG}} & \text{gravity-dominant regime} \\ m=m_c \Leftrightarrow \overline{f_{Q}}=\overline{f_{QG}} & \text{transition-regime} \\ m<m_c \Leftrightarrow \overline{f_{Q}}>\overline{f_{QG}}  & \text{quantum-dominant regime}  \end {array} \right.
\end{equation}
This classification helps us understand the possible regimes in this context\cite{RefRS}.\par 
In many cases, it is more appropriate to work with the concept of energy. The relation (\ref{d1}) can also be derived through  energy considerations. The  quantum Hamilton-Jacobi equation is as follows:
\begin{equation}\label{ham}
-E+\frac{(\nabla S(\mathbf{x},t))^2}{2 m}+U_{\text{cl}}+Q=0.
\end{equation}
In the transition regime, where the averages of quantum force and quantum-gravitational force are balanced, we expect to have a stationary state for which the width of the wave packet is independent of time and the action or principal function, $S$,  only depends on time. In other words, for a stationary state that can be represented in the form $\psi=\psi_0(\mathbf{x}) e^{\frac{iS(t)}{\hbar}}$, the term $\nabla S(\mathbf{x},t)$ in the Hamilton-Jacobi equation (\ref{ham}) vanishes. So, we have:
\begin{equation}\label{hamave}
\overline{E}=\overline{U_{\text{QG}}}+\overline{Q}= \int_{0}^{\infty} \varrho(r) U_{QG} 4\pi r^2 dr+\int_{0}^{\infty} \varrho(r) Q 4\pi r^2 dr,
\end{equation}
which leads to
\begin{equation}\label{hamave2}
\overline{E}= \frac{G \,m^{2}}{2 \sqrt{\pi}\, \sigma_{0}}-\frac{\sqrt{2}G m^{2}}{\sqrt{\pi}\, \sigma_{0}} + \frac{3 \hslash^{2}}{8 m \sigma_{0}^{2}}.
\end{equation}
Now, by minimizing this equation i.e., $\delta \overline{E}=0$, we obtain $\sigma_0 \approx \frac{\hbar^2}{G m^3}$, which is the characteristic width of the wave packet for the transition regime. In this case, the energy of the particle is given by
\begin{equation}\label{sta}
\overline{E}_{\text{stationary}}\approx -\frac{G^2 m^5}{\hbar^2},
\end{equation}
which its negative sign indicates a bound system.
 To obtain the critical values of an object, we shall also use the method of minimizing the Hamilton-Jacobi equation.
\subsection{The Transition Condition for an Object}
For an object, we assume that the only degree of freedom is the coordinates of its center of mass. Thus, the associated wave packet is described by the same wave packet as a point particle. From fundamental physics, we know that for a homogeneous spherical mass distribution with radius $R$ and mass $m$, the classical gravitational potential is given by:
\begin{equation}\label{og}
U_{\text{cl}}(r)=-\frac{ Gm^2}{R} \left(\frac{3}{2}-\frac{1}{2}\frac{r^2}{R^2}\right).
\end{equation}
Now, we obtain $U_{\text{QG}}$ to estimate the transition width for the object. The quantum gravitational energy, which is defined through the relation (\ref{self}), becomes:
\begin{equation}\label{selfo}
\begin{split}
U_{\text{QG}}(r)=&\int_{0}^{r}  -\frac{ Gm^2}{R} \left(\frac{3}{2}-\frac{1}{2}\frac{r^{\prime^2}}{R^2}\right)  \varrho(r^{\prime}) 4\pi r^{\prime^2} dr^{\prime}=\frac{3 m^{2} \sqrt{2}\, G \,{\mathrm e}^{-\frac{r^{2}}{2 \sigma_{0}^{2}}} r}{2 \sqrt{\pi}\, \sigma_{0} R}\\
&-\frac{m^{2} \sqrt{2}\, G \,r^{3} {\mathrm e}^{-\frac{r^{2}}{2 \sigma_{0}^{2}}}}{2 \sqrt{\pi}\, \sigma_{0} R^{3}}
-\frac{3 m^{2} \sqrt{2}\, G \sigma_{0} r \,{\mathrm e}^{-\frac{r^{2}}{2 \sigma_{0}^{2}}}}{2 \sqrt{\pi}\, R^{3}}\\
&-\frac{3 m^{2} G \,\mathrm{erf}(\frac{\sqrt{2}\, r}{2 \sigma_{0}})}{2 R}+\frac{3 m^{2} G \sigma_{0}^{2} \mathrm{erf}(\frac{\sqrt{2}\, r}{2 \sigma_{0}})}{2 R^{3}}.
\end{split}
\end{equation}
Now, as we did for a point particle, we minimize the average energy of the object. The average of the quantum gravitational energy of the object is as follows:
\begin{equation}\label{QGave}
\overline{U_{QG}}=\int_{0}^{\infty} \varrho(r) U_{QG} 4\pi r^2 dr=-\frac{3 G m^{2}}{4 R}+\frac{ G m^{2} \sigma_{0}^{2}}{ R^{3}}\Big(\frac{3}{4}-\frac{1}{\pi}\Big),
\end{equation}
where we have taken the integral over the whole space due to the nature of the Gaussian distribution. Here, the average quantum potential of an object is the same as the average quantum potential of a particle. So, the average of total energy of the object when we consider a stationary wave packet for the transition regime is:
\begin{equation}\label{ham3}
\overline{E}=\overline{U_{QG}}+\overline{Q}=-\frac{3 G m^{2}}{4 R}+\frac{ G m^{2} \sigma_{0}^{2}}{ R^{3}}\Big(\frac{3}{4}-\frac{1}{\pi}\Big)+\frac{3 \hslash^{2}}{8 m \sigma_{0}^{2}}.
\end{equation}
Now, $\delta \overline{E}=0$ gives
\begin{equation}\label{co}
\Sigma_0\approx \left(\frac{\hbar^2}{Gm^3}\right)^{\frac{1}{4}} R^{\frac{3}{4}}= \sigma_0^{\frac{1}{4}} R^{\frac{3}{4}}.
\end{equation}
Here, $\Sigma_0$ denotes the transition width for the object and $\sigma_0$ stands for the transition width of the point particle. This result agrees with the findings of previous studies conducted within the framework of standard quantum mechanics \cite{RefD2}.\par 
If we obtain the critical width through the equality of the averages of the quantum gravitational and quantum forces, we can consider two limiting cases in this problem.
The quantum gravitational force is as follows:
\begin{equation}
f_{\text{QG}}=-\frac{dU_{\text{QG}}}{dr}=\frac{3 G \,m^{2} \sqrt{2}\, r^{2} {\mathrm e}^{-\frac{r^{2}}{2 \sigma_0^{2}}}}{2 \sigma_0^{3} R \sqrt{\pi}}-\frac{G \,m^{2} \sqrt{2}\, r^{4} {\mathrm e}^{-\frac{r^{2}}{2 \sigma_0^{2}}}}{2 \sigma_0^{3} R^{3} \sqrt{\pi}}
\end{equation}
The average of quantum gravitational force is given by
\begin{equation}
\overline{f_{\text{QG}}}=\int_{0}^{\infty} \varrho(r) f_{\text{QG}} 4\pi r^2 dr.
\end{equation}
For a microscopic object, $\sigma_0 \gg R$, so the asymptotic result is
\begin{equation}\label{ftiny}
\Big\vert\overline{f_{\text{QG}}^{(\text{micro})}}\Big\vert=\frac{9 G m^2}{8 \sqrt{\pi}\sigma_0 R}.
\end{equation}
This relation gives a more reliable result when the quantum behavior is significant. In this situation, the wave packet is wide, and the object can be found outside the region $0\leq r \leq R$ most of the time.\par 
For a macroscopic object, $\sigma_0 \ll R$ and we get the following result,
\begin{equation}\label{a3}
\Big\vert\overline{f_{\text{QG}}^{(\text{macro})}}\Big\vert=\frac{15}{16 \sqrt{\pi}}\frac{Gm^2}{R^3}\sigma_0
\end{equation}
which is comparable to the relation of the gravitational force exerted on a particle within a homogeneous mass distribution. For a macroscopic object, we expect the object to spend most of its time within the interval $0\leq r \leq R$.\par 
For a macroscopic object, we equate the average quantum gravitational force in relation (\ref{a3}) with the average quantum force to obtain the critical width of the wave packet. The result is the same as we saw in relation (\ref{co}).
For objects with quantum-dominant behavior, such as a proton, the quantum dispersion is significant. By equating the average quantum gravitational force in relation (\ref{ftiny}) with the average quantum force, we obtain the critical width of the wave packet. This leads to:
\begin{equation}\label{widthl}
\Sigma_0\approx \Big(\frac{\hbar^2}{Gm^3}\Big)^{\frac{1}{2}}R^{\frac{1}{2}}= \sigma_0^{\frac{1}{2}}R^{\frac{1}{2}}.
\end{equation}
This relation gives the transition width $\Sigma_0 \approx 10^6 \text{cm}$ for a proton, which is an acceptable value \cite{RefBassi}.\par 
When $\sigma_0=R$, each of the relations (\ref{co}) or (\ref{widthl}) give 
\begin{equation}\label{R}
\Sigma_0 \approx R=\sigma_0=\frac{\hbar^2}{Gm^3}
\end{equation}
which can be used for intermediate situations. In this situation, the average of quantum gravitational force is as follows:
\begin{equation}
\Big\vert\overline{f_{\text{QG}}^{(\text{intermediate})}}\Big\vert\approx  \frac{Gm^2}{R^2}= \frac{Gm^2}{\sigma_0^2}.
\end{equation}
In the next section, we discuss the meaning of the reduction time based on the quantum theory of motion.
\section{The Reduction Time } \label{sec:3}
We work in the gravity-dominant regime, where only quantum gravitational force is considered. The particle or object falls within its own quantum distribution $\varrho=\psi^*\psi=\mathcal{R}^2$, in configuration space. We can define the time needed for the particle to fall from an initial distance to the center of distribution. We shall see that the quantum motion of the particle or object is oscillatory in this regime, and the reduction time is proportional to one-quarter of the period of oscillation in the short-time estimation. The reduction time is also obtained through the uncertainty in the particle's or object's quantum gravitational energy.
\subsection{The Reduction Time for a Point Particle}
The equation of motion of the particle is as follows:
\begin{equation}\label{b1}
f_{\text{QG}}=-\sqrt{\frac{2}{\pi}}\frac{Gm^2}{\sigma^3_0}r e^{-\frac{r^2}{2\sigma^2_0}}=m\ddot{r},
\end{equation}
where, we have used the relation (\ref{sg}).
By assuming the initial conditions $\dot{r}(0)=0$ and $r(0)=\sigma_0$, and some approximations, it can be shown that the position of the particle obeys the equation
\begin{equation}\label{cos}
r(t)=\sigma_0 \cos \left[\sqrt{2}\left(\frac{2}{\pi}\right)^{\frac{1}{4}}\left(\frac{Gm}{\sigma_0^3}\right)^{\frac{1}{2}}t \right]. 
\end{equation}
For details, See \cite{RefRGG3}.
This shows that the particle has an oscillatory motion through  its probability distribution, with the angular frequency
\begin{equation}
\omega=\sqrt{2}\left(\frac{2}{\pi}\right)^{\frac{1}{4}}\left(\frac{Gm}{\sigma_0^3}\right)^{\frac{1}{2}} \approx \left(\frac{Gm}{\sigma_0^3}\right)^{\frac{1}{2}}
\end{equation}
and the period
\begin{equation}\label{per}
T=\frac{2 \pi}{\omega}=2^{\frac{1}{4}} \pi^{\frac{5}{4}} \Big(\frac{\sigma_0^3}{Gm}\Big)^{\frac{1}{2}} \approx \Big(\frac{\sigma_0^3}{Gm}\Big)^{\frac{1}{2}}.
\end{equation}
The time needed for the particle to travel from the initial distance $r(0)=\sigma_0$  to $r(\tau)=0$ is proportional to the reduction time, which is one-quarter of the period of motion\footnote{In fact, it can be shown that the falling time is obtained through the relation $T_{\text{falling}}\approx \Big(\frac{\sigma_0^3}{Gm}\Big)^{\frac{1}{2}} F\Big(\frac{r_f}{r_i}\Big)$, where $F$ is the elliptical function. Here, $r_i$ and $r_f$ denote the initial and final distances respectively. For details, See Ref.\cite{RefRGG3}}. It can be shown that relation (\ref{per}) is proportional to the reduction time in relation (\ref{karol2}). 
We can simply state that the reduction time of the wave function is proportional to the period of oscillation.  According to  relation (\ref{per}), we conclude that by increasing the mass of the particle for the fixed $\sigma_0$, the period of oscillation decreases and the particle reaches the center of distribution faster. Additionally, the motion is harmonic, but it is not simple harmonic oscillation because the period of motion depends on the initial position of the particle or the amplitude of the oscillation. \par
\begin{figure}[h!] 
\centerline{\includegraphics[width=10cm]{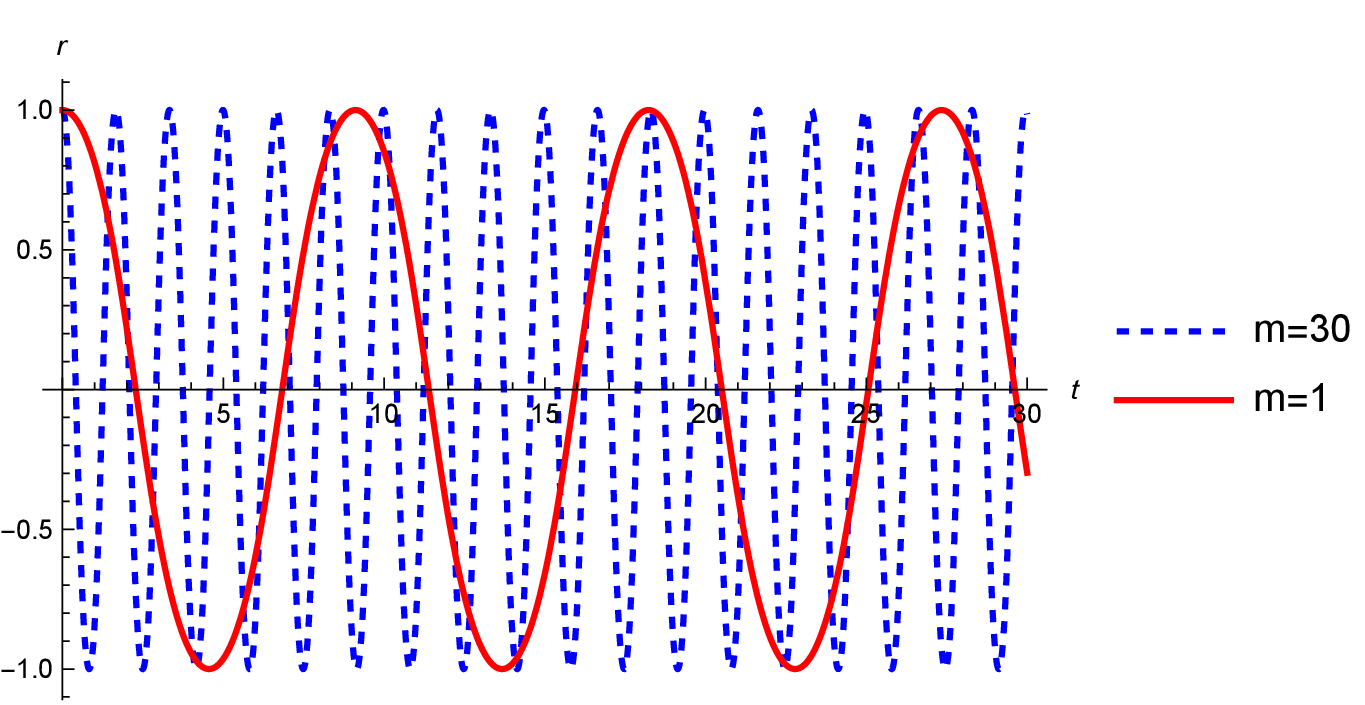}}
\caption{The oscillatory motion of the particle through its quantum distribution in configuration space with initial condition $r(0)= \sigma_0 =1$, in a short-time estimation. The solid  and dashed lines correspond to the values $m=1$ and $m=30$, respectively. Parameters are set to $\hbar=G=1$. The period of motion depends on the mass of the particle. \label{fig:2}}
\end{figure}

\begin{figure}[h!] 
\centerline{\includegraphics[width=10cm]{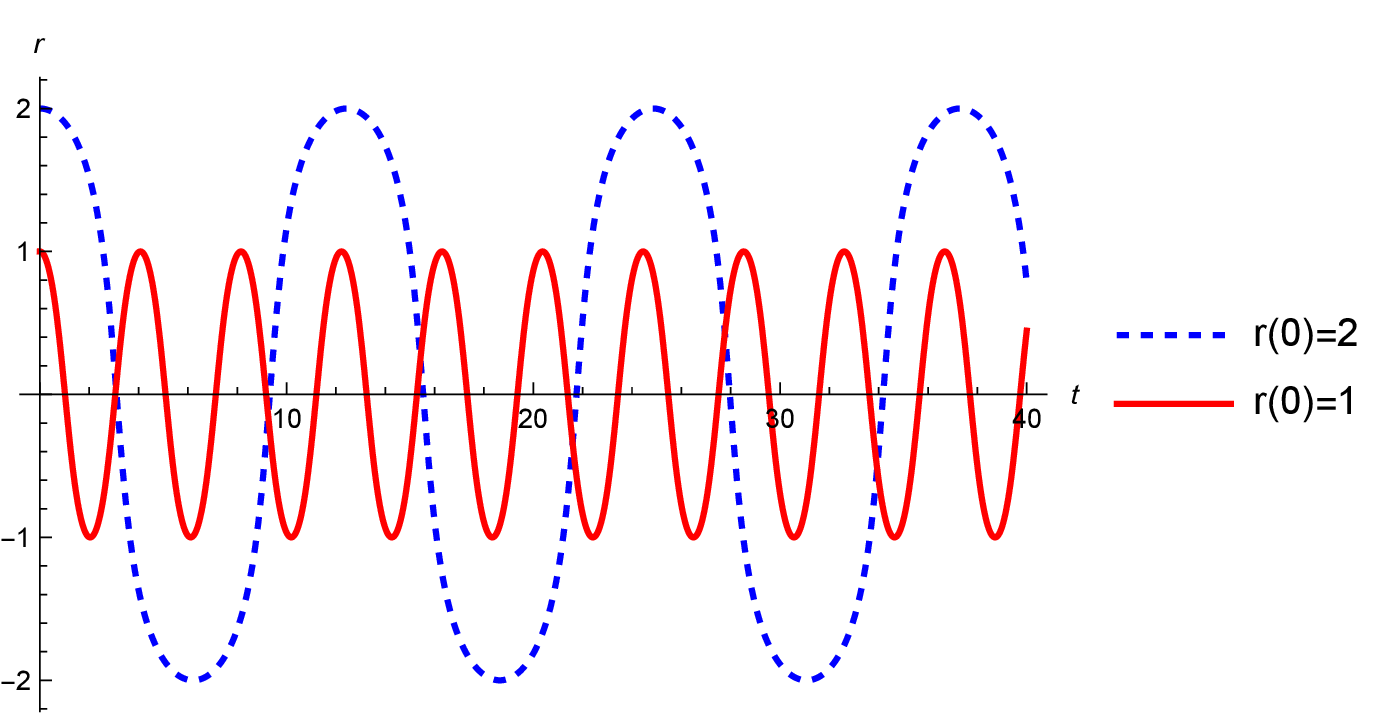}}
\caption{The oscillatory motion of the particle through its quantum distribution in configuration space is not a simple harmonic motion. For both diagrams, $m=5$. For the dashed line, $r(0)=2$, while for the solid line, $r(0)=1$. Parameters are set to $\hbar=G=1$. The period of motion depends on the amplitude of motion. \label{fig:3}}
\end{figure}

 
Equation (\ref{b1}) can be solved numerically to confirm the oscillatory motion of the particle. The results are illustrated in Fig. (\ref{fig:2}) for some arbitrary values of the involved parameters.
The reduction of the wave function does not mean that the particle remains at one point after the collapse of its wave function. Rather, the particle oscillates within an interval that is inversely related to its mass. This interval is so small for a classical particle that we confidently say the particle has a definite location. On the other hand, for a microscopic particle, this interval can be such that the oscillation period lasts for thousands of years, and the uncertainty in the location of the particle is significant.
Figure (\ref{fig:3}), clearly shows that the period of motion depends on the amplitude of motion. The figure has been plotted for two different initial positions of the particle.
By using the relation (\ref{cos}), the quantum gravitational acceleration is obtained as follows:
\begin{equation}\label{tb}
\ddot{r}(t)=g_{\text{QG}}(t)=-2 \sqrt{\frac{2}{\pi}} \frac{G m}{\sigma_0^2} \cos \left[\sqrt{2}\left(\frac{2}{\pi}\right)^{\frac{1}{4}}\left(\frac{Gm}{\sigma_0^3}\right)^{\frac{1}{2}}t \right]
\end{equation}
It is not difficult task to show that the time average of the quantum gravitational acceleration over a period of oscillation is
\begin{equation}
\overline{g_{\text{QG}}} \approx \frac{Gm}{\sigma_0^2}.
\end{equation}
According to these results and arguments, we have:
\begin{equation}\label{time1}
\tau \approx \Big(\frac{\sigma_0^3}{Gm}\Big)^{\frac{1}{2}}=\Big(\frac{\sigma_0}{\overline{g_{\text{QG}}}}\Big)^{\frac{1}{2}}.
\end{equation}
Since we have considered that $\sigma(t) \approx \sigma_0$ during the reduction time, we can substitute 
$\sigma_0=\frac{\hbar^2}{Gm^3}$ into  relation (\ref{time1}) to obtain:
\begin{equation}\label{time2}
\tau=\frac{\hbar^3}{G^2m^5}
\end{equation}
as a completely objective relation for the reduction time in a short-time estimation.\par 
It can be interesting to investigate the particle motion in a mixed regime where the quantum force and the quantum gravitational force are present together. The equation of motion in a short-time estimation is as follows:
\begin{equation}\label{mixed}
m\ddot{r}=\frac{\hbar^2}{4 m \sigma_0^2} r -\sqrt{\frac{2}{\pi}}\frac{Gm^2}{\sigma^3_0}r e^{-\frac{r^2}{2\sigma^2_0}}.
\end{equation}
\begin{figure}[h!] 
\centerline{\includegraphics[width=10cm]{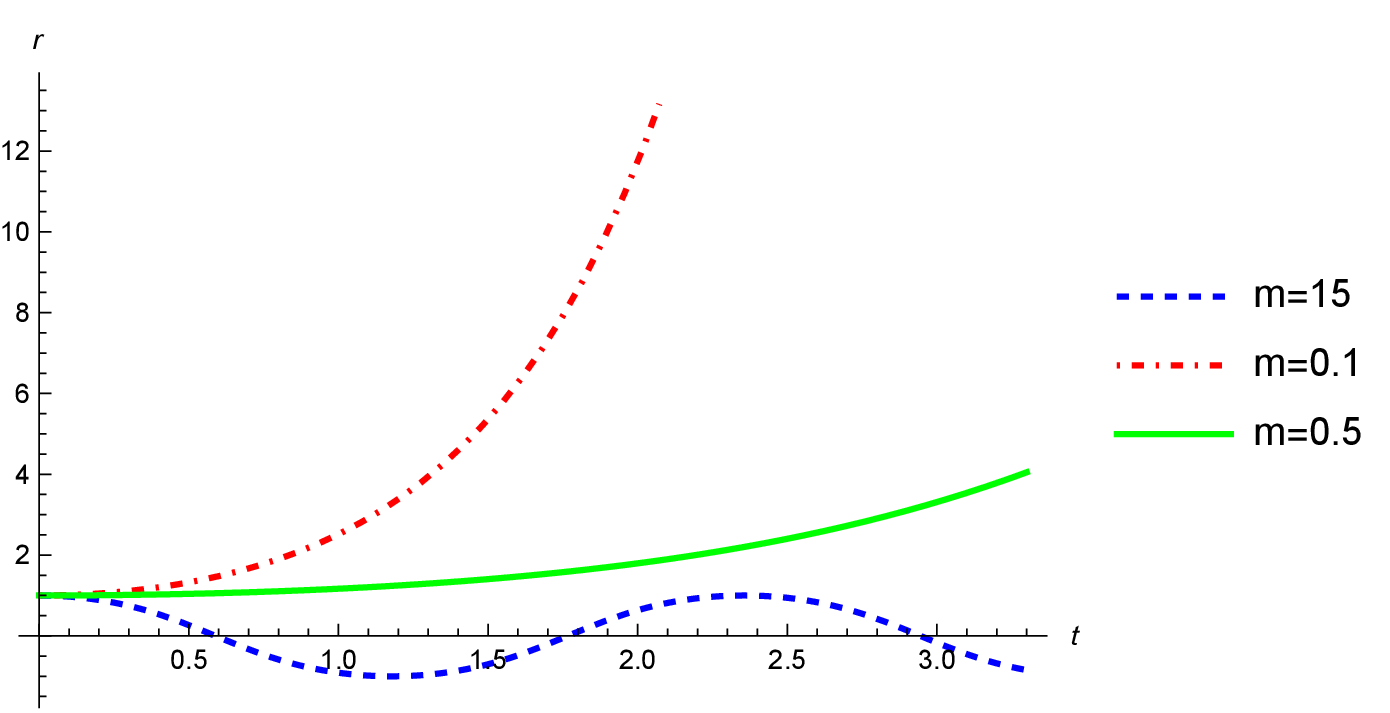}}
\caption{Motion of the particle through its quantum distribution for various mass values in the mixed regime. Initial condition are $r(0)=1$and $\dot{r}(0)=0$, with $\hbar=1$, $G=1$, and initial width $\sigma_0=1$. \label{fig:4}}
\end{figure}
Figure (\ref{fig:4}) shows that for a fixed initial width of the wave packet, by increasing the mass, the particle motion becomes oscillatory. However, by decreasing the mass, the quantum force becomes dominant and the particle moves away from the center of the probability distribution. Thus, we cannot observe the wave function reduction for light or elementary particles unless their wave function is reduced due to the measurement process. The measurement effect can also be related to the gravity caused by the measuring apparatus. Since the wave function of the measuring apparatus is entangled with the wave function of the quantum system, and the measuring apparatus is a macroscopic object, by reducing its wave function, the wave function of the quantum system will inevitably be reduced. This concept is well explained in Ref. \cite{RefP1}.\par 
One can see the whole story as follows:\\
If, for a fixed probability distribution, the mass of the particle is less than the critical mass required to keep the probability distribution stationary, the particle will move away from the origin of the distribution under the effect of the quantum force and the corresponding wave function will never reduce. For a mass equal to the critical mass, the particle should form a stable system.
Now, consider a fixed stationary quantum distribution. A particle with a mass greater than the critical mass falls within its quantum distribution. In fact, it has an oscillatory motion which is not simple. The collapse time is proportional to the period of oscillation. We can also look at the problem in the way that the particle has an uncertainty in its location. More uncertainty means more quantum behavior. If the mass of the particle exceeds the critical mass, the particle falls towards the center of the probability distribution and performs an oscillatory motion within its quantum probability distribution. In other words, the particle is trapped by gravity. This confinement means that the location of the particle is determined with greater certainty. As the mass increases, classical behavior also increases. This fluctuation also exists for the chair in our room, but it is so small that we know the location of the chair with high certainty. However, the period of such an oscillation for a proton is so long that we have to wait thousands of years to observe the collapse of a proton.\par 
We can also estimate the reduction time through the uncertainty principle. When the particle falls into its own quantum distribution from a distance $\sigma_0$, it  reaches the center of distribution after time $\tau$. The average value of momentum at the time $\tau$, is $p=mu\approx mg\tau$. Now, according to $\Delta r \Delta p \sim \hbar$, we get $mg\tau \sigma_0 \sim \hbar$. So, we will have
\begin{equation}\label{tun1}
\tau =\frac{\hbar}{mg\sigma_0},
\end{equation}
where, the denominator of the relation is proportional to the average of quantum gravitational energy of the particle between $0\leq r \leq \sigma_0$. Thus, we can suggest the general relation
\begin{equation}\label{gt}
\tau=\frac{\hbar}{\Delta}=\frac{\hbar}{\vert U_{\text{QG}}(\sigma_0)-U_{\text{QG}}(0) \vert} 
\end{equation}
which for a point particle gives
\begin{equation}
\tau=\left(\frac{\sigma_0^3}{Gm}\right)^{\frac{1}{2}} \approx \frac{\hbar^3}{G^2	m^5}.
\end{equation}
This is the same as relation (\ref{time2}). We shall use this method in obtaining the reduction time for an object.
\subsection{The Reduction Time for an Object}
\begin{figure}[h!] 
\centerline{\includegraphics[width=12cm]{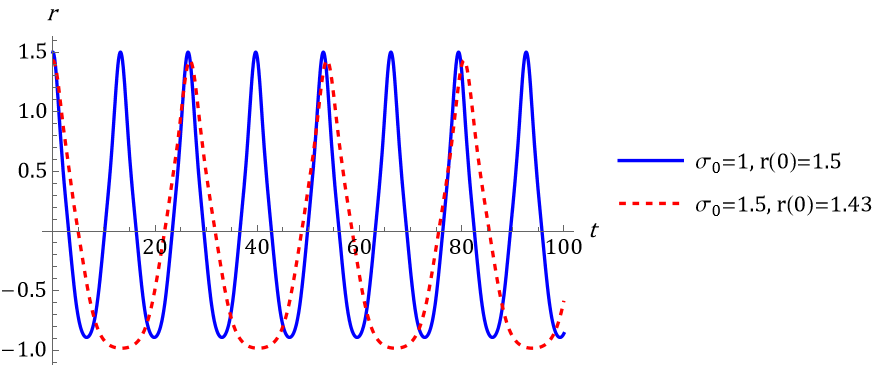}}
\caption{The motion of the object within its quantum distribution in the gravity-dominant regime. Parameters are set to $R=1$, $m=1$, $\hbar=1$, and $G=1$.
For the dashed line, $\sigma_0=1.5$ and $r(0)=1.43$; for the solid line, $\sigma_0=1$ and $r(0)=1.5$. The period of oscillation is dependent on the initial position of the object and the width of its associated wave packet. \label{fig:5}}
\end{figure}

We realized that for a point particle, the reduction time is proportional to the period of oscillation with good approximation, and the associated relation (\ref{time1}) confirms the previous results obtained in the framework of standard quantum mechanics. If we want to do the same for an object, we find that the corresponding equation of motion is more complicated, and its solution cannot be obtained analytically unless we use some approximations. Therefore, we cannot easily derive an analytical formula for the period of motion. However, the equation of motion can be solved numerically, and the period of oscillation can be evaluated. To observe the oscillatory motion of the object in the gravity-dominant regime, consider the equation of motion as follows:
\begin{equation}\label{objectmotion}
m \ddot{r}=\left(\frac{3}{2}\sqrt{\frac{2}{\pi}} \frac{G m^2}{\sigma_0^3 R} r^2 -\frac{1}{2} \sqrt{\frac{2}{\pi}} \frac{G m^2}{\sigma_0^3 R^3} r^4\right) \exp \Big(-\frac{r^2}{2 \sigma_0^2}\Big).
\end{equation}
The solution of this equation for different values of parameters is seen in Fig. (\ref{fig:5}). The figure shows that the period of motion depends on the initial position of the object. Also, the amplitude of motion is not the same above and below the horizontal axis because the first term in the parentheses of relation (\ref{objectmotion}) acts as a repulsive force.  
Naturally, this is due to the mass distribution and has nothing to do with the nature of the gravitational force. We did not have such a term for a point particle. For various initial conditions, equation (\ref{objectmotion}) can be solved to give us an estimation for the reduction time, which is proportional to the period of oscillation. \par
While solving the above equation provides the desired solutions for objects of varying sizes and masses and allows us to estimate the  reduction time, it is intriguing that by considering the uncertainty in the particle's quantum gravitational energy, we also derive an approximated analytical relation for the reduction time.
When, the object's center of mass is at $r=\sigma_0$ in configuration space, an uncertainty in quantum gravitational energy can be attributed to this distribution. In fact we consider the uncertainty in quantum gravitational energy during the reduction time $\tau$, when the object's center of mass has reached $r=0$. So, we use the relation (\ref{gt}) to get:
\begin{equation}\label{ttt2}
\tau=\frac{\hbar}{\Delta}=\hbar\Big\vert \alpha \frac{Gm^2 \sigma_0^2}{R^3}  -\beta \frac{Gm^2}{R} \Big\vert^{-1}.
\end{equation}
where the values of $\alpha$ and $\beta$ depend on the degrees of approximation.  
For a ball with the size $R=5$ cm, the mass $m=100 g$ and $\sigma_0\approx 10^{-12} cm$, the reduction time is on the  order of  $10^{-23} s$. The value of $\sigma_0$ has been  estimated using relation (\ref{co}). In fact, we estimate the critical width $\Sigma_0$ for an object.
For a flea egg with a size of approximately $0.5 mm$ and a mass of about $10^{-5}g$, the reduction time is around $10^{-11} s$.
\par 
As relation (\ref{widthl}) was obtained for a micro object by considering the asymptotic limit of the  quantum gravitational force ($\sigma_0 \gg R$), the asymptotic form of the quantum gravitational energy should be considered here to provide a reliable result for the reduction time. For a micro object, the leading order of the asymptotic relation ($\sigma_0 \gg R$) of the quantum gravitational energy is as follows:
\begin{equation}\label{UQGa}
U_{\text{QG}}^{(\text{asympt})}=-\frac{2\sqrt{2}}{5\sqrt{\pi}}\frac{Gm^2 r^3}{R \sigma_0^3},
\end{equation}
which is obtained through relation (\ref{selfo}). 
Now, substituting this relation into the relation (\ref{gt}), leads to
\begin{equation}\label{tmicun}
\tau=\frac{5}{4}\frac{\sqrt{2\pi}\hbar R}{Gm^2}\approx \frac{\hbar R}{G m^2}.
\end{equation}
For a proton, this gives the value $\tau \sim 10^{15}s $ , which indicates that we do not see its collapse normally. This result is also in agreement with the results obtained in the framework of standard quantum mechanics. In Bohmian quantum mechanics, the position of a particle or object is a hidden variable, meaning we do not know which trajectory is chosen during its motion. However, when the particle moves along any trajectory within the ensemble, it must obey the equation (\ref{vel1}). Assuming the initial position of the particle or object is near the initial wave packet width, we find that the approximated analytical results align with those obtained from standard quantum mechanics. However, the deterministic structure of Bohmian quantum mechanics, which accounts for underlying processes, allows us to solve the problem for any initial position and velocity derived from the initial wave packet. 
\begin{figure}[h!] 
\centerline{\includegraphics[width=12cm]{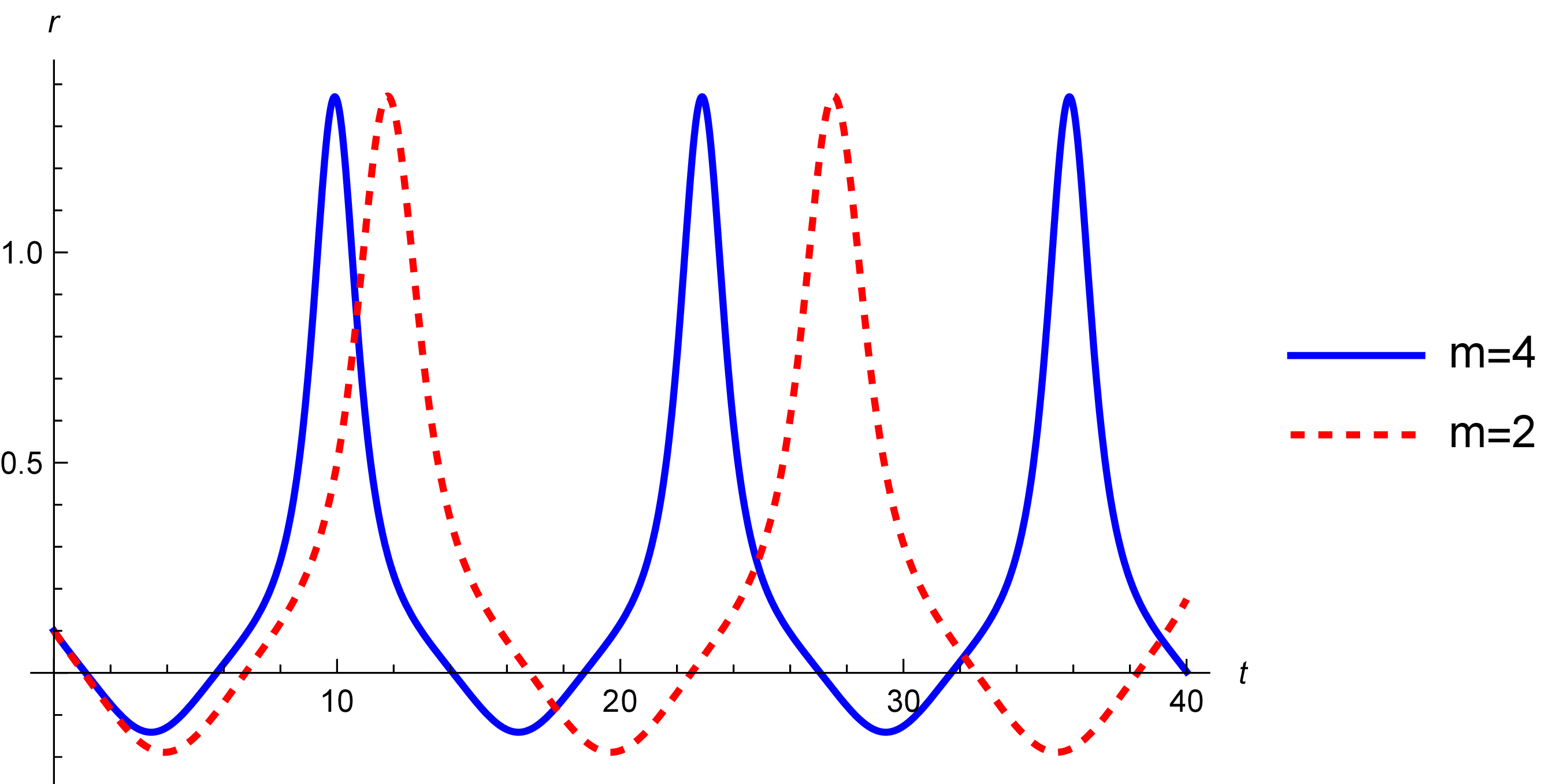}}
\caption{The motion of the object within its quantum distribution in the gravity-dominant regime for different values of mass. Parameters are set to $R=1$, $\sigma_0=1$, $\hbar=1$, and $G=1$. The initial position and velocity are $r(0)=0.1$ and $r^{\prime}(0)=-0.1$ respectively.
For the dashed line, $m=2$; for the solid line, $m=4$. The period of oscillation is inversely dependent on the mass of the object. \label{fig:6}}
\end{figure}
In Fig. \ref{fig:6}, we assume the object has an initial velocity towards the center of the distribution and examine how the period of motion changes with variations in mass, while keeping the wave packet width and the size of the object constant. As expected, increasing the mass decreases the period of motion, allowing the object to reach the center of the distribution more quickly.\par 
According to these investigations and results, the problem of gravitational reduction of the wave function within the framework of Bohmian quantum mechanics appears to yield reliable results. Consequently, this approach merits further attention.

\section{Conclusion}
A Bohmian perspective for gravity-induced wave function reduction was provided. Through this approach, it is possible to investigate the problem of gravitational reduction of the wave function with concrete concepts such as force and particle motion. Examining the deviation between nearby trajectories in the ensemble led to the condition of equilibrium between quantum and quantum gravitational forces, which is used to obtain the critical width for the transition between the quantum and classical worlds.
We also obtained the relation (\ref{cri}), which classifies the different regimes of motion in terms of the ratio of particle mass to the critical mass.
In a gravity-dominant regime, where only the quantum gravitational force is considered, we investigated the equation of motion of the particle and found that the particle performs oscillatory motion within a certain range of its quantum distribution. What is discussed as the collapse time of the wave function in standard quantum mechanics is related here to the period of the particle's oscillatory motion.
We showed that this oscillatory motion is not simple and depends on the amplitude of the motion. In other words, particles with wider wave packets take longer to reach the center of the probability distribution and vice versa.
The behavior of particle motion was illustrated in Figs. (\ref{fig:2}) and (\ref{fig:3}).
Additionally, the motion of the particle was examined in a mixed regime, where both quantum and gravitational forces are present. It was observed that as the mass decreases, the particle moves away from the distribution center due to the quantum force, and the reduction of the corresponding wave function is never observed. Conversely, as the mass of the particle increases, the gravitational regime prevails, and the particle performs oscillatory motion within a range determined by its mass.
 \par
For an object, we obtained the critical width using the concepts of Bohmian quantum mechanics. The results are in agreement with those obtained by Diosi in her studies. We showed that for micro objects, one must consider the asymptotic form of the quantum gravitational force ($\sigma_0\gg R$) to achieve the correct result for the critical width of the associated wave packet.
The equation of motion of an object (Eq. \ref{objectmotion}) was investigated numerically, revealing a non-simple oscillatory motion in a gravity-dominant regime. Using equation (\ref{objectmotion}), it is possible to check the motion of the object for different initial conditions and parameter values. While the period of motion can be obtained numerically through the equation of motion, we estimated the reduction time through the uncertainty in the quantum gravitational energy of the object to achieve approximate analytical results.
This approach provides a clearer understanding of the underlying processes, thanks to the deterministic structure of Bohmian quantum mechanics. It seems that this approach can be further studied to obtain more comprehensive results.


\end{document}